\documentstyle[12pt]{article}
\begin{document}
\thispagestyle{empty}

\begin{center}
\LARGE \tt \bf {Gravitational stability of inflaton and torsion in Einstein-Cartan-Klein-Gordon cosmology with kinky potentials}
\end{center}

\vspace{2.5cm}

\begin{center} {\large 

L.C. Garcia de Andrade\footnote{Departamento de
F\'{\i}sica Te\'{o}rica - Instituto de F\'{\i}sica - UERJ

Rua S\~{a}o Fco. Xavier 524, Rio de Janeiro, RJ

Maracan\~{a}, CEP:20550-003 , Brasil.E-mail:garcia@dft.if.uerj.br}}
\end{center}
\vspace{2cm}

\begin{abstract}
Gravitational stability of torsion and inflaton field in a four-dimensional spacetime de Sitter solution in scalar-tensor cosmology where Cartan torsion propagates is investigated in detail. Inflaton and torsion evolution equations are derived by making use of a Lagrangean method. Stable and unstable modes for torsion and inflatons are found to be dependent of the background torsion and inflaton fields. Present astrophysical observations favour a stable mode for torsion since this would explain why no relic torsion imprint has been found on the Cosmic Background Radiation in the universe. 
\end{abstract}

\newpage
\section{Introduction}
\indent
 Some attempts have been made recently \cite{1,2,3,4} to investigate the cosmological perturbations  \cite{5} in spacetimes with torsion. The simplest example is the Capozziello and Stornaiolo  \cite{1} which considered gauge invariant approach by simply considered the extension of general relativistic small perturbations and substituting into it the effective matter density and pressure into the evolution equations where the effective quantities bring the spin-torsion density on the expressions. Earlier Piskareva and Obukhov \cite{2} have considered the vector (rotational), scalar (matter density) and tensorial (gravitational waves) small perturbations of spinning fluid matter without nevertheless without torsion. More recently Garcia de Andrade \cite{3} has considered the stability of inflatons torsion and metric fluctuations around de Sitter solutions in Einstein-Cartan-Klein-Gordon (ECKG) cosmology. In this Letter we follow the approach considered by Maroto and Shapiro \cite{4} who investigated the existence and stability of de Sitter inflationary solutions for the string-inspired fourth-derivative gravity theories with torsion and dilatons in arbitrary $D-dimensional$ spacetime. One of the main features of their work as far as the study of the role of torsion is concerned is that Cartan torsion field is nondynamic in the sense that torsion obeys an algebraic equation and not a differential equation. On the other hand in their theory de Sitter solution fluctuates to a more general solution of the string action. In this Letter we propose to work with the Lagrangean method to investigate a scalar-tensor action where now the Lagrangean would depend also on torsion and not only on the metric cosmic scale factor and inflaton field. The present letter is organized as follows. In the section $2$ we present the derivation of the field equations from the action of this scalar-tensor cosmology. In section $3$ we investigate the fluctuations on the inflaton and torsion and derive the evolution equations for both fields. We also show by solving this equation in a simple case that torsion and inflatons possess stable and unstable modes depending on the relation between torsion and inflaton backgrounds. Our solutions and equations are clearly distinct to the ones obtained previously by Maroto and Shapiro in reference $2$. Section $4$ presents some brief discussion.
Of course during the paper we consider inflatons instead of dilatons since we are dealing with four-dimensional inflationary cosmology and not higher-dimensional string theories. It is also important to note that here we do not attempt to find an exct solution of the ECKG field equations of gravity \cite{5} but only to investigate the behaviour and stability of inflatons and torsion against the background de Sitter solution, this is possible since we have shown very recently \cite{3} that ECKG equation is reduced to Einstein-de Sitter equation around the inflaton constant background field ${\phi}_{0}$.

\section{Inflaton and Torsion Equations in Inflationary Cosmology.}
\indent
In this section we present the torsion and inflaton sections derived from the action \cite{4}

\begin{equation}
S_{M} = \int{d^{4}x \sqrt{g}[ R({\Gamma})(1+k{\phi}^{2}) + 4({\partial}{\phi})^{2}-V({\phi})]}
\label{1}
\end{equation}
where $ g $ is the determinant of the de Sitter metric and $R({\Gamma})$ is the non-Riemannian Ricci scalar given by

\begin{equation}
R = -6 H^{2}_{0}- \dot{T}- 6 T^{2}(t)
\label{2}
\end{equation}
where ${\phi}$ is the inflaton potential $T(t)$ is the torsion zero component of the torsion vector while $H_{0}$ is the de Sitter expansion factor which reads
\begin{equation}
ds^{2} = dt^{2} - e^{2H_{0}t}(dx^{2}+dy^{2}+dz^{2})
\label{3}
\end{equation}
Therefore the above action reads
\begin{equation}
S_{M} = \int{d^{4}x  e^{3H_{0}t}[- 6(H^{2}_{0}+\frac{1}{6}\dot{T}+{T}^{2})(1+k{\phi}^{2}) + 4{\dot{\phi}}^{2}-V({\phi})]}
\label{4}
\end{equation}
where here we choose the inflaton potential $V({\phi})= \frac{\lambda}{4}{\phi}^{4}$ kinky potential frequently used in inflationary cosmology. Writing this action in the form 
\begin{equation}
S = \int{dt e^{3H_{0}t} L(T,\dot{T},{\phi},\dot{\phi})}
\label{5}
\end{equation}

here $L$ represents the Lagrangean function. Variation of this action leads to the following field equations for torsion and inflaton
\begin{equation}
3H_{0}\frac{{\partial}{L}}{{\partial}{\dot{T}}}+ \frac{d}{dt}\frac{{\partial}L}{{\partial}{\dot{T}}}-\frac{{\partial}{L}}{{\partial}{T}}=0
\label{6}
\end{equation}
By the symmetry of the Lagrangean the expression for the inflaton field is given by

\begin{equation}
3H_{0}\frac{{\partial}{L}}{{\partial}{\dot{\phi}}}+ \frac{d}{dt}\frac{{\partial}L}{{\partial}{\dot{\phi}}}-\frac{{\partial}{L}}{{\partial}{\phi}}=0
\label{7}
\end{equation}
Substitution of Lagrangean L in equation (\ref{4}) one obtains 
\begin{equation}
8\ddot{\phi}+24H_{0}\dot{\phi}+2k[6H_{0}^{2}+\dot{T}+6{T}^{2}]-4{\phi}^{3}=0
\label{8}
\end{equation}
This equation has a simple solution with the ansatz ${\phi}=e^{{\alpha}t}$. This ansatz reduces the inflaton equation to an algebraic equation 
\begin{equation}
{\alpha}^{2}+24H_{0}{\alpha}-2kR=0
\label{9}
\end{equation}
After some algebra substitution of expression 
\begin{equation}
{\phi} \rightarrow {\phi}_{0}+ {\delta}{\phi}
\label{10}
\end{equation}
into equation for the evolution of inflaton yields
\begin{equation}
{\delta}\ddot{\phi}+ 3H_{0}{\delta}\dot{\phi}+\frac{3}{2}{\alpha}_{0}{\phi}_{0}+\frac{k}{4}{\phi}_{0}[{\delta}{\dot{T}}+12k{\phi}_{0}{\delta}T]=0
\label{11}
\end{equation}
Before complete this evolution equation we need to compute torsion fluctuation equation. This yields 
\begin{equation}
T(t)= \frac{1}{4}(1+k{\phi}^{2})^{-1}+\frac{1}{12}\frac{d}{dt}ln(1+k{\phi}^{2})
\label{12}
\end{equation}
Let us now considered the fluctuation of the Cartan torsion against a constant background torsion $T_{0}$ given by
\begin{equation}
T \rightarrow T_{0}+ {\delta}T
\label{13}
\end{equation}
Substitution of this expression into the equation (\ref{11}) yields in the first order perturbation the expression
\begin{equation}
{\delta}T(t)= \frac{1}{12}{{\phi}_{0}}^{2}\delta{\dot{\phi}}-\frac{1}{2}k{\phi}_{0}{\delta}{\phi}
\label{14}
\end{equation}
while the zero-$0$ order perturbation leads to the constraint on torsion background $T_{0}$ 
\begin{equation}
T_{0}(t)= \frac{1}{4}{{\phi}_{0}}^{2}
\label{15}
\end{equation}
where we also consider the small perturbation of the inflaton field above and drop second-order perturbations like $({\delta}{\phi})^{2}$. Now substitution of ${\delta}T$ into the expression  (\ref{11}) yields
\begin{equation}
a{\delta}\ddot{\phi}+ b{\delta}\dot{\phi}+c{\delta}{\phi}+d=0
\label{16}
\end{equation}
where the constants ${a,b,c,d}$ are connected to the background quantities and are explicitly given by
\begin{equation}
a=(1+\frac{k^{2}{{\phi}_{0}}^{2}}{48})
\label{17}
\end{equation}
\begin{equation}
b=[3H_{0}-\frac{k^{2}}{8}{{\phi}_{0}}^{2}]
\label{18}
\end{equation}
\begin{equation}
c=\frac{3}{2}{\alpha}_{0}
\label{19}
\end{equation}
\begin{equation}
d={\phi}_{0}c
\label{20}
\end{equation}
where ${\alpha}_{0}=[k(H_{0}^{2}+T^{2}_{0}]-\frac{\lambda}{4}{\phi}_{0}^{2}]$.Equation (\ref{16}) can be considered as a simple linear differential equation with constant coefficients which can be easily solvable to yield the solution
\begin{equation}
{\delta}{\phi}(t)= -{\phi}_{0}+ c_{\pm} e^{{\beta}_{\pm}t}
\label{21}
\end{equation}
where ${\beta}_{\pm}$ are two constants given by
\begin{equation}
{\beta}_{\pm}=[-\frac{1}{2}\frac{(b{\pm}\sqrt{b^{2}-4ac})}{a}]
\label{22}
\end{equation}
To ilustrate our procedure let us consider the special case to simplify matters where ${\alpha}_{0}$ vanishes which yield the relation a fourth order algebraic equation to determine the background inflaton field
\begin{equation}
{\phi}^{4}_{0}-\frac{\lambda}{k}{\phi}_{0}^{2}+16H_{0}^{2}=0
\label{23}
\end{equation}
This equation can be solve by the MAPLE V algebraic computation program. Since it is a fourth-order algebraic equation it yields the following four roots
\begin{equation}
{{\phi}_{0}}^{1,2}={\pm}\frac{1}{2}\sqrt{2}\sqrt{k({\lambda}+\sqrt{({\lambda}^{2}-64k^{2}H_{0}^{2})})}
\label{24}
\end{equation}
\begin{equation}
{{\phi}_{0}}^{3,4}={\pm}\frac{1}{2}\sqrt{-2k(-{\lambda}+\sqrt{({\lambda}^{2}-64k^{2}H_{0}^{2})})}
\label{25}
\end{equation}
Now with ${\alpha}_{0}=0$ one is able to simplify the solution to
\begin{equation}
{\delta}{\phi}(t)= -{\phi}_{0}+ c_{0} e^{-[3H_{0}-\frac{k^{2}}{8}{\phi}_{0}^{2}]t}
\label{26}
\end{equation}
Note that this solution will possess a stable or unstable carachter depending upon the relation between the Hubble parameter $H_{0}$ and the inflaton background field and the nonminimal coupling constant. It is easy to show that slow roll approximation \cite{6} applied on the equation (\ref{16}) yields the following expression 
\begin{equation}
b{\delta}\dot{\phi}+c{\delta}{\phi}+d=0
\label{27}
\end{equation}
This important special case leads to the following simple solution 
\begin{equation}
{\delta}{\phi}= -\frac{d}{a} + e^{-\frac{c}{b}t}
\label{28}
\end{equation}
which explicitly yields
\begin{equation}
{\delta}{\phi}= -{\phi}_{0} + e^{-\frac{c}{b}t}
\label{29}
\end{equation}
It is easy to show that even if the inflaton background vanishes the inflaton fluctuation will be stable. Of course the stability also depends upon the background torsion since they are related as we saw earlier in this letter. A whole family of solutions is found out in the paper. Second-order perturbations equations are extremely complicated and should be addressed on a separated paper.
 
\section*{Acknowledgements}
I would like to express my gratitude to Professors Yuri Obukhov and Ilya Shapiro for helpful discussions and sugestions on the subject of this paper. Financial support from CNPq. is gratefully acknowledged.

\end{document}